\documentclass[aps,prl,floatfix,onecolumn,tightenlines]{revtex4}

\usepackage{amsmath}
\usepackage{latexsym}
\usepackage{graphicx}
\usepackage{color}
\usepackage{fancyhdr}
\usepackage{graphicx}
\usepackage{subfigure}

    \include{Abstract}

\begin{document}

\date{September 2015}

\title{Part I - A case study in post-secondary mathematics: the importance of mental health awareness }

\author{Janelle Resch\footnote{This paper was edited and written with the help of Eric Ocelewski (iocelews@uwaterloo.ca).}}
\affiliation{jresch@uwaterloo.ca}

\begin{abstract}
\noindent Due to the shift of civilization from the Industrial Age to the Information Age, mathematical literacy has become a necessity in the twenty-first century. However, in order to learn and contribute to the mathematical community, one has to be in a state of good mental health. Where `good mental health,' is defined as one who has developed a set of healthy coping strategies, while being in a positive learning environment, and having a social support system. Traditionally, universities have been the venue of such higher learning. If these institutions want to remain as thriving grounds for higher education and engines of research, post-secondary institutions need to be aware of these factors and actively contribute to the well-being of its students and faculty. Unfortunately, students do not always receive the support necessary to be mentally healthy. The purpose of this paper is to examine how social awareness and sensitivity of mental health in a university setting is a key component for individuals to flourish academically and grow personally. Several voluntary surveys completed by first and second year math students at the University of Waterloo will be presented. The surveys investigated the students' experiences at the university, particularly in the Mathematics Department. In addition, this paper explores an alternative way to structure math classes, specifically assignment scheduling, in order to help students develop healthy study habits. Instead of giving students weekly-assignments which is typically done at the University of Waterloo for the core math classes, mini-assignments were assigned after each lecture. This experiment was conducted for the second year linear algebra class, and the two methods are compared and statistically analyzed. The group who completed mini-assignments after each class consisted of 459 students, whereas the group that completed weekly-assignments consisted of 387 students. The results indicate that the more frequent assignment schedule helped students increase confidence and overall grades while reducing anxiety and stress. Finally, this paper briefly discusses the importance of socialization, specifically for young mathematicians and scientists, and potential consequences of reduced social exposure in the Digital Age.  
\end{abstract}

\maketitle

\noindent \textbf{Keywords:} Mental health, mathematics education, STEM, University of Waterloo, student surveys, Ebbinghaus' curve of forgetting, assignment schedule, simple random sample, box plots, histograms, t-tests, F-tests, z-tests, ANOVA tests, amygdala, autism-spectrum disorder.      
\noindent\section{Introduction}

According to the Mood Disorder Society of Canada (MDSC), \textit{mental health} and \textit{mental illness} is defined as follows:

\begin{itemize}
\item[] \textbf{Mental Health:}
	
\textit{(or well-being) is an ideal we all strive for. It is a balance of mental, emotional, physical and spiritual health. Caring relationships, a place to call home, a supportive community, and work and leisure all contribute to mental health. However, no one's life is perfect, so mental health is also about learning the coping skills to deal with life's ups and downs the best we can.}

\item[] \textbf{Mental Illness:	}

\textit{is a serious disturbance in thoughts, feelings and perceptions that is severe enough to affect day-to-day functioning} \cite{1}.

\end{itemize}
\noindent However, in this paper, I will instead be using the term \textit{poor mental health} rather than \textit{mental illness} to avoid the negative association with the term `illness.' 

Some specific types of poor mental health that one typically hears about are for instance: schizophrenia, depression, bi-polar disorder, anxiety disorders, eating disorders, borderline personality disorder, autism-spectrum disorder, post-traumatic stress disorder (PTSD), etc. \cite{1}. Why such forms of poor mental health develop depends on many factors such as genetics and one's environment, but this is an over simplification of the issue \cite{2}. The issue is highly individual, and globally complex as a result. Researchers still know very little about poor mental health and the relation between one's environment and gene expression. It is a topic that is slowly starting to be taken more seriously, especially given recent media coverage which describes poor mental health as an increasing pandemic \cite{3}. Nonetheless, there is still significant and substantial stigma, and naive notions when it come to mental health.

The younger generations are especially vulnerable to poor mental health. According to the MDCS, 18$\%$ of Canadians alone between the ages of 15 - 24 are reported to have poor mental health or substance abuse problems. Sadly, suicide accounts for 24$\%$ of all deaths among Canadians aged 15 - 24, and 16$\%$ of all deaths for the age group 25 - 44. The people with the highest rate of depression are people under the age of 20; and people between the ages of 20 - 29 have the highest rate of anxiety \cite{1}. This is a major problem, especially if we want to have a functional, altruistic society in the future. This is the generation that we expect will be solving some, if not all, of the world's most difficult problems. Issues such as climate change, population growth and population sustainability, economic inequality (perhaps even the collapse of the monetary system), the impact of integration of automation technologies and robots into everyday society, and so on \cite{3}, \cite{4}, \cite{5}.

In order to solve these problems, people need to be scientifically literate and mathematically competent and proficient. In general, modern functioning society is highly dependent on technology; particularly on science, technology, engineering and mathematics (STEM) educated individuals. With the high probability of most repetitive jobs vanishing due to technological developments in automation \cite{4} -  STEM education and training will be essential and necessary for twenty-first century careers. To learn such material requires a lot of intense, focused and consistent work. To complete such work, stable and reliable mental health must exist. This implies an individual's mental health is just as important as physical health for such intense learning to take place. Hence, it is essential that we address poor mental health. This is particularly important for our young people who grew up with the Internet due to their increased physical isolation relative to previous generations. Unfortunately, very little research has been done on the impact of mental stability and the development of young adults' learning capacities and motivations.

As a young mathematics instructor at a post-secondary institution, the University of Waterloo in Canada, mental health became a major concern of mine, consuming substantial time.  The longer I teach these young people, the more I see their struggle in learning due to poor mental health \cite{6}. Unfortunately, most of the time the problem is not recognized or taken seriously. Observing this first hands motivated me to write this paper to help the well-being of my students. The focus will be to present supporting evidence that a shift in the primary and post-secondary education systems' culture is needed to be:

\begin{itemize}
\item more aware of mental health,
\item aware of how poor mental health affects the learning process, and
\item aware of how an institution's awareness affects their students.
\end{itemize}

\noindent\section{A commentary on surveys}

In order to gain more perspective and information on the mental health and opinions of my students, I created several surveys for them to complete if so desired. At the start of the Fall 2014 term at the University of Waterloo, I made the surveys available to the students I had taught in the previous terms, and the students I was teaching at the time. More specifically, I prepared these surveys using Google Forms for the mathematics students at the University of Waterloo. This included students in the undeclared math program (first and second year students), as well as students in the applied math, pure math, statistics, combinatorics and optimization, mathematical physics, and double degree math-business programs. The students who completed these surveys did so out of voluntary compliance; it was not part of any class and the students were informed the results would be summarized in a paper. Although I was skeptical that anyone would complete them, to my surprise, many students were willing and enthusiastic about completing (and helping create) the surveys. This demonstrated to me that the students wanted their opinions to be known; they just required an appropriate outlet. 

For the survey results shown below, the total number of students who answered `yes' or `no' is given. The number of students who did not answer `yes' or `no' but rather left a comment, is not included in the mentioned numerical values. However, some of these comments can be found in Appendix B.

\noindent\section{Data From Surveys}

The purpose of the initial surveys was to determine the students' views on mental health and how the university setting influences their mental well-being. The following results were obtained:

\begin{itemize}
\item 82$\%$ of 238 students said: \textit{the professor I have for a class influences how much I work in that class.}

\item 88$\%$ of 236 students said: \textit{I am more motivated in class if I like the professor.}

\item 66$\%$ of 227 students said: \textit{a professor's awareness on mental health influences my enjoyment in a class.}

\item 95$\%$ of 234 students said: \textit{professor's should be aware of mental health.}

\item 84$\%$ of 221 students said: \textit{discussions on mental health are important.}

\item 77$\%$ of 215 students said: \textit{I believe I am aware of my own mental health.}
 
\item 77$\%$ of 212 students said: \textit{I believe that my peers and I are not aware of each other's mental health.} 

\item 65$\%$ of 136 students said: \textit{we did not have discussions on mental health in high school.} 40 \% of the remaining students (35$\%$ of 136 students) who said: \textit{we did have discussion on mental health in high school}, also said: \textit{I do not think the school did a good job.}
 
\item 88$\%$ of 102 students said: \textit{professors should be able to take mental health into consideration.}

\item 61$\%$ of 102 students said: \textit{mental health has not been addressed thoroughly at university.}

\item 72$\%$ of 107 students said: \textit{I think professors should have training on mental health.}

\item 48$\%$ of 167 students said: \textit{I think that the professors at the University of Waterloo are not aware of mental health.}
\end{itemize}

The following questions were also asked on the student surveys:
\begin{itemize}
\item[] Q1: Are there any departments at the university that you feel are less aware of mental health?
\item[] Q2: Do you think poor mental health will become a larger issue in society as time progresses?
\item[] Q3: Has the university affected your mental health?
\item[] Q4: Do you think mental health is compromising your success in classes?
\end{itemize}

Fifty-two students responded to Q1, 58 responded to Q2, 83 answered Q3 and 102 answered Q4. The distribution of student responses to Q1 can be seen in Fig. 1. It can be seen that 52$\%$ of students said the Mathematics Department in particular was not very aware of the students' mental health. Engineering was the second highest ranked department at 17$\%$.

\begin{figure} [ht]
\centering
\includegraphics[scale=0.45]{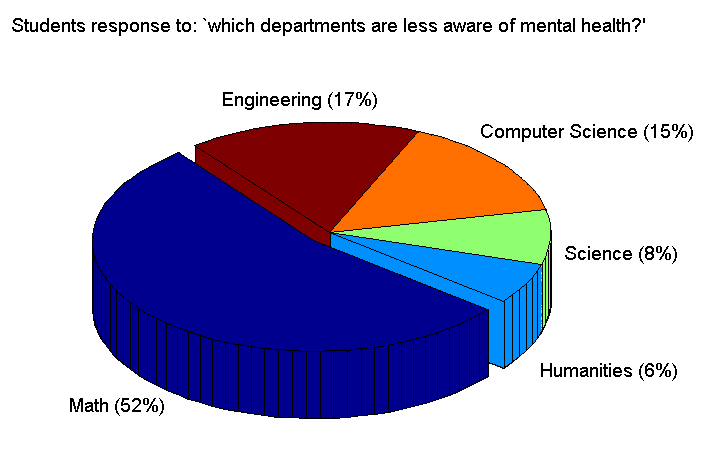}
\caption[]{Distribution of the 52 student responses to Q1.}
\label{fi:1}
\end{figure}

\begin{figure} [ht]
\centering
\includegraphics[scale=0.74]{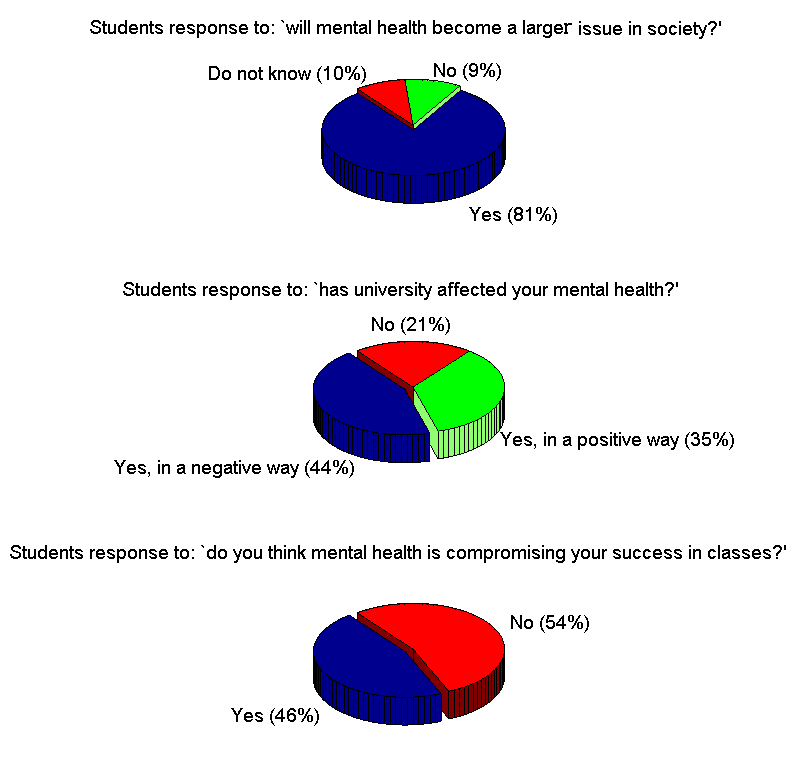}
\caption[]{Distribution of student responses to: Q2 top (58 responses), Q3 middle (83 responses), Q4 bottom (102 responses).}
\label{fi:1}
\end{figure}

The distribution of student responses to Q2, Q3 and Q4 can be seen in the top, middle and bottom plot of Fig. 2, respectively. It was found that 81$\%$ of students thought that poor mental health will become a larger issue in society as time naturally progresses. This is a hypothesis of mine, and my conjecture is that poor mental health is related to the larger amounts of information that everyone is exposed to on a regular basis. Consider that within a single minute, more information is uploaded onto the Internet than a person 200 year ago was exposed to in their entire lifetime \cite{7}. Such volumes of information will have some of influence on our minds and consciousness. Humanity has progressed this far due to natural selection and thus, how our physical bodies have evolved. Surely our minds and consciousness will evolve as our environment changes with respect to time. Therefore, I hypothesize that `mental illness' is an evolution of our minds responding to the volume of information exposed to us. 

In the middle plot of Fig. 2, one can see that 44$\%$ of students surveyed think the university has influenced their mental health in a negative way; whereas 35$\%$ say the university has influenced their mental health in a positive way. Unfortunately, as depicted in the bottom plot of Fig. 2, 46$\%$ of the students surveyed think their success in the classroom is being compromised because of their mental health. Although the potential connection to gender (and the corresponding issues) will not be discussed in this paper (it will be the focus of a future paper), I hypothesize that gender equality (or a lack thereof) also has an effect on the mental health of women at the university. I speculate this is an issue because of my own experiences and from the stories women at the university have shared with me.

\noindent \section{Restructuring of Assignments}


\noindent \subsection{Introduction and Purpose}

I always tell my students that learning mathematics is like learning a language; and no one is born speaking automatically. A person needs a certain amount of exposure and practice before being able to speak any language. Therefore, it cannot be expected for individuals to form proper sentences before becoming familiar with words, sentence structure and grammar. It is natural and in fact, a part of the learning process to mispronounce words or use them incorrectly. This same argument also holds for the language of mathematics. People can find learning mathematics to be challenging and even overwhelming because of the unfamiliarity. Making mistakes is an unavoidable part of the learning process. However, this does not imply that mistakes (i.e., failing) should be treated as catastrophes unless one approaches failure in an unproductive way.

\begin{figure} [ht]
\centering
\includegraphics[scale=0.62]{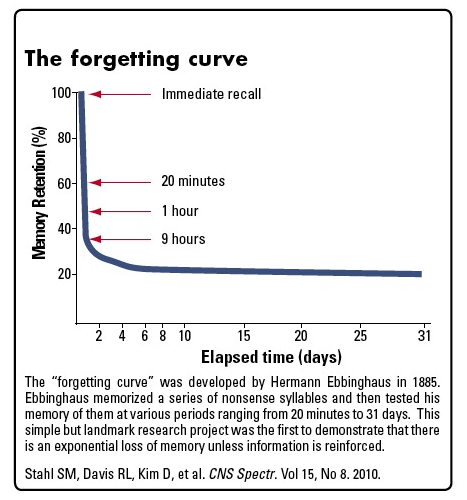}
\caption[]{Ebbinghaus forgetting curve based on fading theory \cite{11}.}
\label{fi:1}
\end{figure}
\vspace{2mm}

Studies have consistently shown that review is essential in order to learn new information. The study of memory was initiated by the German psychologist named Hermann Ebbinghaus. In 1885, he published a paper describing what we now call the \textit{learning curve} and the \textit{curve of forgetting} \cite{12}, \cite{13}. Ebbinghaus' curve of forgetting is depicted in Fig. 3. It illustrates the exponential decay of learning new information from a single learning session \cite{13}. It indicates that if one does not \textit{actively} review newly learned information within the first 24 hours, only 33$\%$ of the information originally absorbed will be retained. Therefore, students should review their notes within the first 24 hours after attending class to help store the information to long-term memory. Fig. 4 depicts how reviewing on a regular basis can help ensure one does not forget new material \cite{14}. However, it is essential to review from day one. If students consistently prepare for class, learning will be a much more rewarding and enjoyable experience.

\begin{figure} [ht]
\centering
\includegraphics[scale=0.72]{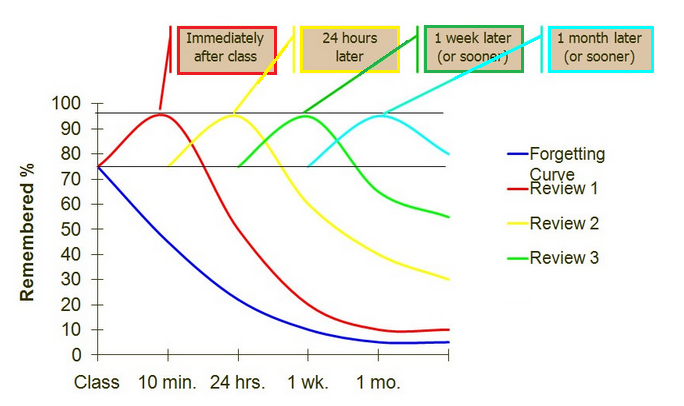}
\caption[]{Ebbinghaus forgetting curve and importance of reviewing \cite{14}.}
\label{fi:1}
\end{figure}

Due to these findings, I structured my classes for the last couple of years to have \textit{end of lecture} (EoL) assignments rather than the typical weekly assignments given in university math classes. However, this was a decision that I personally made for my own sections. I wanted to investigate if such a structure would be beneficial in a multi-section course, i.e., if all sections did EoL assignments rather than weekly assignments. In the Fall 2014 term at the University of Waterloo, I had the opportunity to carry out such an experiment. I coordinated the second year honours math linear algebra class, referred to as \textit{MATH 235}. This course is required for all math majors, but typically is not considered the most enjoyable course.

\noindent \subsection{Feedback from Students}

For the MATH 235 Fall 2014 class, all 459 students had a class every Monday, Wednesday and Friday. After each class, the students were given three to five questions that were due on the next day of the class. Upon surveying the students three weeks into the term, the following was discovered:

\begin{itemize}
\item 94$\%$ of 110 students said: \textit{I think the EoL assignments are making me keep up with the course material.}

\item 91$\%$ of 110 students said: \textit{I feel like the EoL assignments are improving my learning.}

\item 60$\%$ of 102 students said: \textit{I feel less anxious coming to class after doing an EoL assignment.}

\item 52$\%$ of 98 students said: \textit{I look forward to coming to class more after completing an EoL assignment.}
\end{itemize}

\noindent After the midterm test the students said the following:

\begin{itemize}

\item 79$\%$ of 159 students said: \textit{I think the EoL assignments helped me be more prepared for the midterm.} 

18$\%$ of the 159 students said: \textit{the EoL assignments did not make a difference [in preparing them for the midterm].}

\item 54$\%$ of 142 students said: \textit{I feel like the EoL assignments influenced my confidence for the midterm in a positive way.}

25$\%$ of the 142 student said: \textit{the EoL assignment did not influence [my confidence for the midterm] at all.}

\item 37$\%$ of 159 students said: \textit{I feel like the EoL assignments helped reduce my test anxiety for the midterm.}

23$\%$ of the 159 students said: \textit{I do not have any sort of test anxiety.} 
\end{itemize}

\noindent After the final exam, the students said the following:

\begin{itemize}
\item 90$\%$ of 119 students said: \textit{I enjoyed the [MATH 235] class.}
 
\item 90$\%$ of 124 students said: \textit{I think the EoL assignments helped me learn and keep up with the material regardless of liking the EoL assignments.}

\item 61$\%$ of 115 students said: \textit{I enjoyed the EoL assignments.}

\item 66$\%$ of 122 students said: \textit{the EoL assignments helped me structure my study habits for MATH 235.}

\item 45$\%$ of 93 students said: \textit{I have applied such study habits to other classes or am planning to do so next term.}

\item 68$\%$ of 130 students said: \textit{I understood more of the material in MATH 235 compared to my other classes.}

27$\%$ of the 130 students said: \textit{there was no difference in understanding compared to my other classes.}

\item 45$\%$ of 109 students said: \textit{the EoL assignments helped my mental well-being.}

27$\%$ of the 109 students said: \textit{ I do not know if the EoL assignments helped my mental well-being.}
\end{itemize}

\noindent \subsection{Descriptive Statistics}

At the end of term, the marks of all 459 students who took and completed MATH 235 in Fall 2014 were analyzed. I hypothesized that the Fall 2014 group would outperform other MATH 235 sections that instead completed weekly assignments rather than EoL assignments. To test against my hypothesis, I also analyzed the marks of 387 students who took and completed MATH 235 in the Fall 2013 and Spring 2014 terms. For all three of these terms I was one of the instructors and the same material was taught. In particular, I had two sections in both fall terms, and one section in the Spring 2014 term. However, one of my sections from Fall 2013 and Spring 2014 were not included in the sample of 387 students. This is due to the fact that my sections completed EoL assignments whereas the other sections had weekly assignments consisting of the same questions. For convenience, I will refer to the Fall 2014 group as the \textit{EoL assignment group} and the remaining group as the \textit{weekly assignment group}.

\begin{table}[h]
\begin{center}
\begin{tabular}{ lc c c c c}
\hline
& Assign. Average & Midterm Mark & Final Mark & Overall Grade \\ \hline 
Mean &72.93005181 &70.39896373 & 59.87823834 & 66.55958549 \\
Standard Error & 1.141249365 &1.002159567 &1.015469067 & 0.915371757 \\
Median & 79 & 71 & 62 & 67\\
Mode & 96 & 75 & 64 & 61 \\
Standard Deviation & 22.42199242 & 19.68931146 & 19.95080164 &17.98420155\\
Sample Variance &502.7457439 & 387.6689859 & 398.0344862 & 323.4315053\\
Kurtosis &1.251196139 & 2.134767813 & 0.636438169 & 0.560222138\\
Skewness &-1.290170072 & -1.078715037 & -0.675483433 & -0.634431278\\
Confidence Level (95$\%$) & 2.243861437 & 1.970390762 & 1.996559165 & 1.799753366\\ \hline \\
\end{tabular}
\caption[]{Descriptive statistics of the weekly assignment group.}
\end{center}
\end{table}

For both groups of students, the \textit{marks composition data set}, i.e., the breakup of the evaluation method, consisted of the students: assignment average, midterm mark, final mark and overall grade in MATH 235. The assignments were cumulatively worth 10$\%$ of the students ' overall grade. The EoL assignment group was assigned 32 assignments throughout the term, whereas the weekly assignment group was given 10 assignments. The midterm and final exam were worth 25$\%$ and 65$\%$ of the overall grade, respectively. The descriptive statistics of the marks data set for the weekly assignment group and EoL assignment group can be seen in Tables 1 and 2, respectively. 

\begin{table}[h]
\begin{center}
\begin{tabular}{ lc c c c c}
\hline
& Assign. Average & Midterm Mark & Final Mark & Overall Grade \\ \hline 
Mean & 68.07259821& 74.46623094 &70.81481481& 71.31633636\\
Standard Error& 0.990440257&0.851981117&0.810359931&0.724626167\\
Median&74.64516129&77&73&73.63225806\\
Mode & 94.38709677&95& 66&74.3483871\\
Standard Deviation&21.21947462&18.25308651&17.36138234&15.52459772\\
Sample Variance&450.2661033&333.1751672&301.4175966&241.0131345\\
Kurtosis&0.62697467&3.030615182&1.631463296&0.715859109\\
Skewness&-1.082880529& -1.391674584&-0.960573545&-0.772909234\\
Confidence Level (95$\%$) &1.946370622&1.674276671&1.592484504&1.424004197\\ \hline \\
\end{tabular}
\caption[]{Descriptive statistics of the EoL assignment group.}
\end{center}
\end{table}

We can see that the weekly assignment group had a higher assignment average than the EoL assignment group. However, the EoL assignment group's mean for the midterm exam, final exam and overall grade was higher than the weekly assignment group. The standard error, median, mode, standard deviation, sample variance, Kurtosis value, skewness and confidence level is also given in the tables for both groups. To be scientific, it is important to point out the possibility that the EoL assignment group could have been a stronger group of students academically. However, such a possibility can only be investigated if more data is collected and analyzed. It is also important to mention that many students (roughly 12 - 40) from both the weekly and EoL assignment groups that were enrolled in other sections attended my lectures or office hours. Additionally, if students from the other sections requested to use my personal notes prepared for my students, they were given access to these resources. Approximately 15$\%$ - 20$\%$ of students from other sections used my personal notes.

\begin{figure}[h]
\centering
\subfigure[Weekly assignment group.]{ 
\includegraphics[scale=0.48]{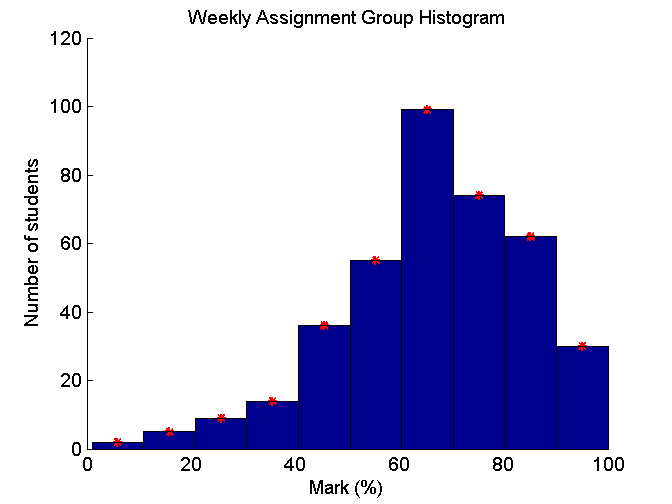}}
\subfigure[EoL assignment group.]{ 
\includegraphics[scale=0.48]{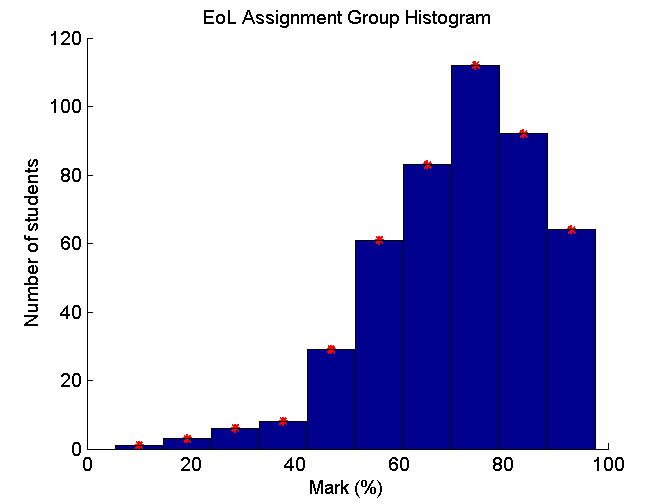}}
\caption[]{Histogram of the students' overall grade for weekly assignment group (left) and EoL assignment group (right).}
\end{figure}

The corresponding histogram of the weekly and EoL assignment groups' overall grades can be seen in Fig. 5a and 5b, respectively. A histogram with a normal distribution fit of both groups' overall grade can be seen in Fig. 6. These figures illustrate that there is a skew to the right for both groups of students. However, the EoL assignment group seems to be more skewed than the weekly assignment group. This comparison is depicted in Fig. 6.

\begin{figure} [ht]
\centering
\includegraphics[scale=0.45]{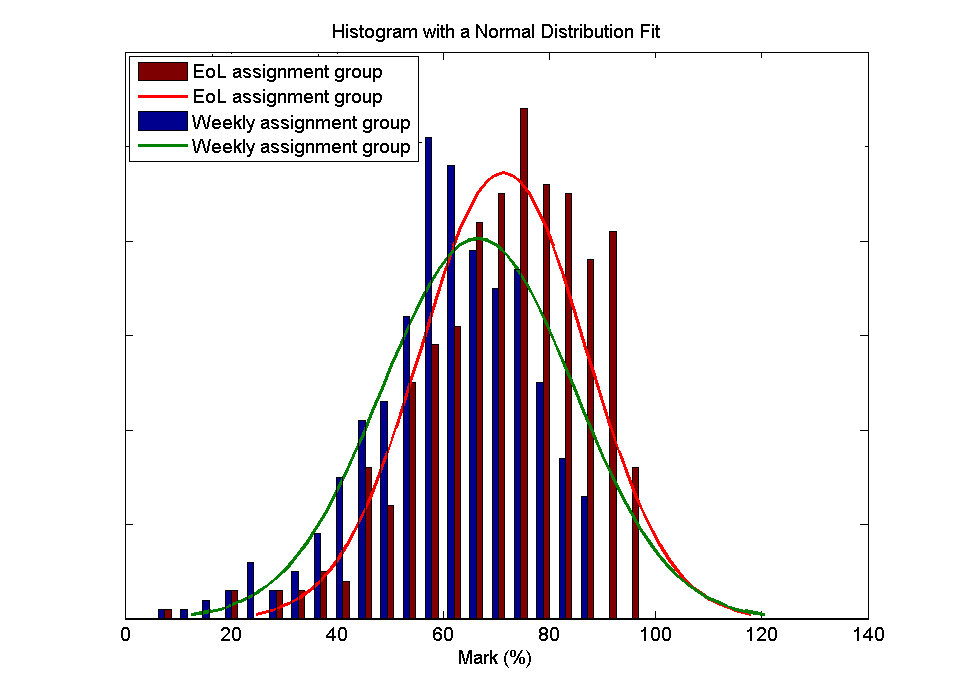}
\caption[]{Histogram with a normal distribution fit of weekly and EoL assignment groups' overall grade data.}
\label{fi:1}
\end{figure}

One can also see a set of four box plots for the weekly assignment group and EoL assignment group in Fig. 7a and 7b, respectively. In each plot, the boxes graphed from left to right represent the students' assignment average, midterm mark, final mark and overall grade obtained in MATH 235, respectively. Notice that the assignment box plots for both groups in Fig. 7 is similar. However, the midterm, final and overall grade box plots and their corresponding relationships to each other differ. In particular, it appears that the EoL assignment group has more consistent, evenly distributed grades compared to the weekly assignment group. The width (i.e., variation of range) of the box plots for the midterm, final and overall grade match each other rather well for the EoL assignment group. Yet, for the weekly assignment group, the ranges of the box plots vary more, especially with respect to each other. Also, the assignment average box plot for the weekly assignment group consists of the highest grades relative to other box plots. However, this is not the case for the EoL assignment group. Since the assignment average box plots for both Fig. 7a and 7b are similar, yet there is more variation between the boxes in Fig. 7a rather than Fig. 7b, this may suggest that the frequency at which the students had to complete assignments in the EoL assignment group allowed the students to retain more information.

\begin{figure}[h]
\centering
\subfigure[Weekly assignment group with sample size of $n=387$.]{ 
\includegraphics[scale=0.48]{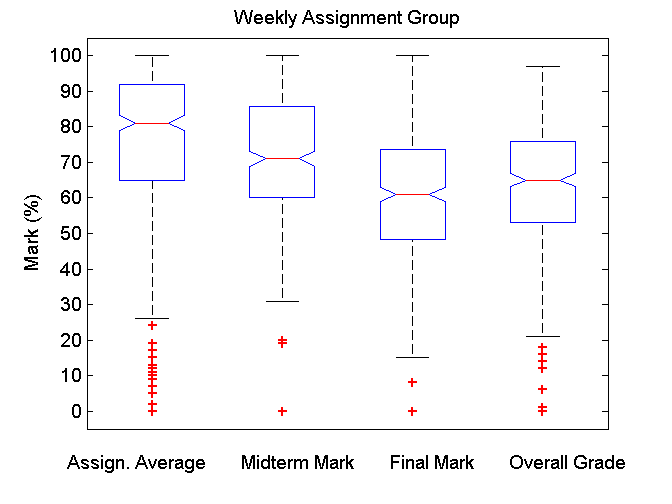}}
\subfigure[EoL assignment group with sample size of $n=459$.]{ 
\includegraphics[scale=0.48]{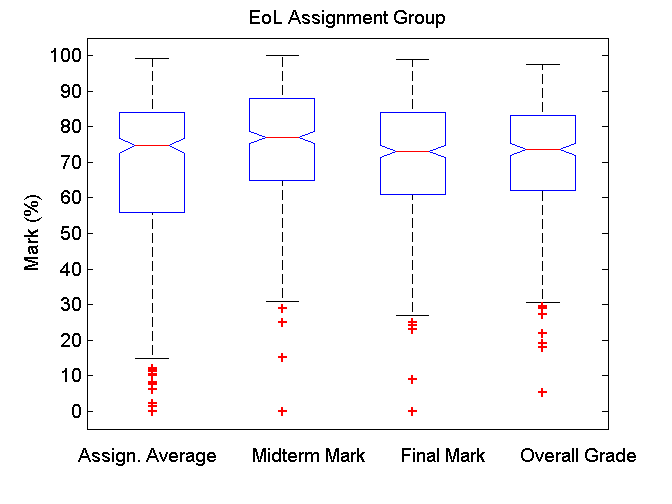}}
\caption[]{Box plots from left to right of students' assignment average, midterm mark, final mark and overall grade.}
\end{figure}

\noindent \subsection{Further Plots}

In Fig. 8, the final mark that each student obtained was compared with their assignment marks. Recall that the weekly assignment group had to complete 10 assignments throughout the term, whereas the EoL assignment group was given 32 assignments to complete. Due to this frequency variation of assignments, Fig. 8a depicts the percentage of assignments the students completed versus the students' overall mark in MATH 235. Fig. 8b plots the students' assignment average against their overall grade in MATH 235. The line of best fit for the two groups in both Fig. 8a and 8b is also depicted. 

In both the left and right plots in Fig. 8, a positive correlation can be seen. In the right plot, the line of best fit for both groups has a similar slope. However, the blue line (representing the EoL assignment group) is shifted by approximately $+10 \%$ relative to the red line (representing the weekly assignment group). In Fig. 8a, the blue line of best fit is similar to the blue line shown in Fig. 8b. However, the slope of the red line of best fit in Fig. 8a is different compared to the slope of the red line in Fig. 8b. In Fig. 8a, the red line is not as steep as the blue line. In particular, the slopes seen in Fig. 8a, are approximately $\frac{1}{3}$ and $\frac{1}{5}$ for the blue and red line, respectively. The blue line is always above the red line after the one intersection point at approximately $(35, \; 50)$. Hence, if a student from the EoL assignment group only completed 35$\%$ of the assignments, they were more likely to obtain a higher grade than a student in the same situation in the weekly assignment group. Note that 35$\%$ assignment completion for the EoL assignment group corresponds to approximately 11 assignments, which is roughly the cardinality of the entire assignment set for the weekly assignment group.

\begin{figure}[h]
\centering
\subfigure[Assignments completed vs overall mark.]{ 
\includegraphics[scale=0.52]{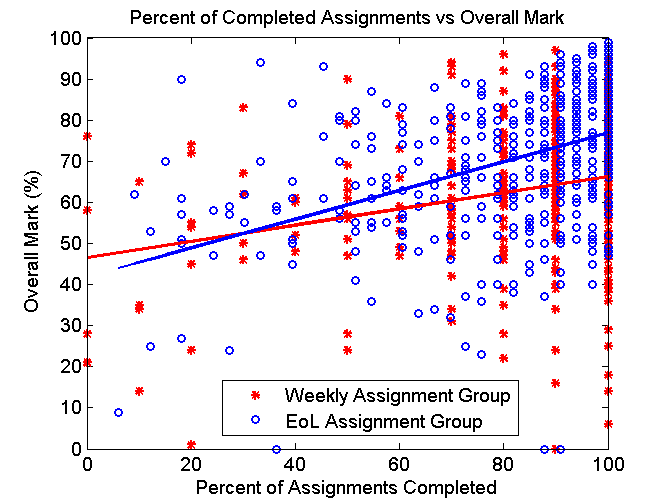}}
\subfigure[Assignment average vs overall mark.]{ 
\includegraphics[scale=0.52]{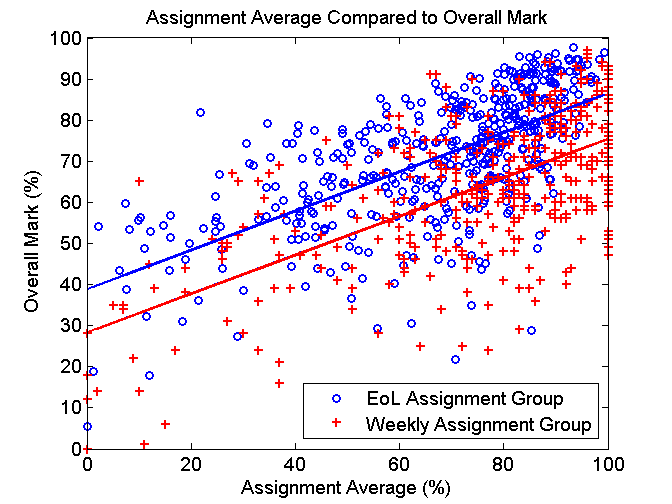}}
\caption[]{Data comparing information about students' assignments and overall mark.}
\end{figure}

Therefore, the data seems to suggest that the higher the frequency of evaluation (not necessarily dependent on marks), the better the measurement of the students' understanding. This demonstrates a decoupling of the assignment average from the overall grade for the EoL assignment group, hence giving a more accurate measure of the students' progress. More precisely, by comparing the information in Fig. 8a and 8b, it appears that the actual assignment grades for the EoL assignment group does not necessarily indicate that the students' understanding is stagnant, or that the numerical value properly characterizes the students' competency. One of the goals of the EoL assignments was for the students to make their mistakes on the assignments rather than on the exams. Thus, even if a student's assignment average is not very high, they had the opportunity to correct and learn from their mistakes at a faster rate due to the higher frequency of evaluation and feedback. This observation also seems consistent from the box plots shown in Fig. 7b.

The final observation from Fig. 8a I will discuss is potentially related to mental health. Notice there are several students who completed less than $50\%$ of their assignments, but still received a decent grade in the course. In particular, there were 38 students in the weekly assignment group and 13 students in the EoL assignment group that completed at most $50\%$ of their assignments but still obtained a final grade of over $65\%$ (which is the average students need to maintain in the Math Department at the University of Waterloo). Interestingly, it appears from Fig. 8a that it is less likely to have such `outliers' in the EoL assignment group. Perhaps this has to do with helping the students structure a study schedule, or giving them the opportunity to seek help more often. Nonetheless, I wanted to examine the number of these `outliers' because several of them (not all) are students with poor mental health. I can say this with confidence because I have spoken to several of these students (also see Appendix B for some commentary from the students). 

Overall, there are a variety of mental health issues or environmental reasons (e.g., domestic violence or sexual violence) that could result in students not attending class or submitting work on time. However, this does not imply that students intend for such things to happen, nor does it imply that the students are not working on their own. Thus, it is very possible to have several outliers, i.e., students who do not submit assignments but who still do well in classes, because they work without focusing on deadlines. For instance, some of the students I spoke with suffer from depression. It is common for students with depression to spend more time in bed, especially during conventional work hours. Recent studies have revealed that ``individuals with depressed mood show as much as a 12$\%$ reduction in memory compared to individuals without depressed mood when depressive thoughts are present, but perform similarly to individuals without depressed mood when depressive thoughts are not present'' \cite{15b}. Therefore, if a student is having a terrible day, it very well may be more productive for the student to attend to their mental health and then work later on their own. That being said, if a student does struggle with mental health and finds that they continually cannot keep up with their studies, such a student should carefully reflect on their course load. It is vital for students to understand their own limitations and adapt new strategies or schedules accordingly. Even if professors do everything in their power to help or accommodate students with poor mental health, it is still up to the student to communicate with their instructors, and take hold of their own responsibilities, lifestyle and needed accommodations. 

\noindent \subsection{Statistical Analysis: Analysis of Variance}

An exhaustive statistical analysis comparing the EoL assignment group and weekly assignment group can be found in Appendix A. It was strongly suggested to me by professional statisticians that considering too many statistical tests is not a good idea because eventually, an issue will arise. I am not sure I agree with such a statement. In my view, if a reasonable data set is collected, it seems sensible to think that the statistical tests would give consistent results. However, I am not a statistician, so perhaps this is a naive notion. Nonetheless, for full disclosure purposes please refer to Appendix A to review all the tests I performed on the data. 

For the statistical tests presented in this section, a simple random sample of 100 individuals from both groups was taken 16 separate times and averaged. I will refer to these new populations as the \textit{random EoL assignment group} and the \textit{random weekly assignment group}. These data sets were also graphed as a set of box plots which can be seen in Fig. 9. The structure of Fig. 9 is similar to Fig. 7. Notice that for each group, the relationship between the box plots depicting the students' assignment average, midterm mark, final mark and overall grade obtained in MATH 235 exacerbates the trends seen in Fig. 7. This observation seems to further support the claims made above while analyzing Figs. 7 and 8.

\begin{figure}[h]
\centering
\subfigure[Randomly sampled weekly assignment group.]{ 
\includegraphics[scale=0.48]{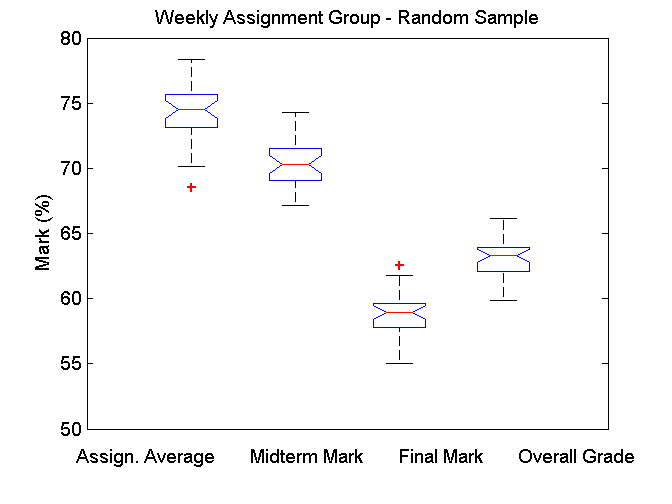}}
\subfigure[Randomly sampled EoL assignment group.]{ 
\includegraphics[scale=0.48]{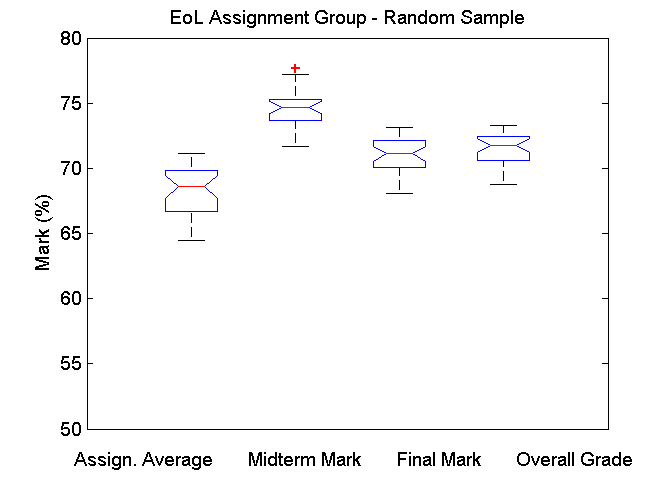}}
\caption[]{Box plots from left to right of students' assignment average, midterm mark, final mark and overall grade.}
\end{figure}

The random weekly and EoL assignment groups' data was also used for further statistical analysis. An analysis of variance (ANOVA) test was used to analyze the statistically significant differences between the two groups. In particular, the one-way ANOVA test was performed on the two mutually independent groups in which the means of the two groups were compared. Thus, the null hypothesis, denoted by $H_0$, was defined as follows:
\begin{itemize}

\item[$H_0:$] there is no statistically significant difference between the randomly sampled populations, against the alternative that there is a statistically significant difference.
\end{itemize}

\noindent The one-way ANOVA test was performed separately for the four different categories of randomized data, i.e., the randomly averaged assignment average, midterm mark, final mark and overall grade data. A test decision for $H_0$ is returned as well as a table with seven columns. The first column shows the source of variability; the second depicts the sum of squares, denoted by SS, due to each source; the third shows the degrees of freedom, denoted by df, associated with each source; the fourth contains the mean squares, denoted by MS, which is the ratio $\frac{SS}{df}$; the fifth shows the F-statistic, denoted by F, which is the ratio of the mean squares; the sixth column contains the p-value, denoted by p; and finally, the seventh column shows the critical F-value, denoted by $F_{\text{crit}}$. If $F>F_{\text{crit}}$, which corresponds to a very small p-value, then the null hypothesis $H_0$, is rejected. The results can be seen below in Table 3. 

\begin{table}[h]
\begin{center}
\begin{tabular}{ lc c c c c c c c}
Results for sampled assignment average data: \\
\hline
Source of Variation & SS & df & MS & F & p & $F_{crit}$ \\ \hline 
Between groups & 216.082 & 1 & 216.082&59.35 & 1.36e-08 & 4.171 \\
Error (within group) &109.229 & 30 & 3.641 & & \\
Total & 325.311& 31 \\ 
\hline \\
Results of sampled midterm mark data. \\
\hline
Source of Variation & SS & df & MS & F & p & $F_{crit}$ \\ \hline 
Between groups & 102.030 & 1 & 102.030 &42.72& 3.15e-07 & 4.171 \\
Error (within group) &71.651& 30 & 2.388 & & \\
Total & 173.682& 31 \\ 
\hline \\
Results of sampled final mark data. \\
\hline
Source of Variation & SS & df & MS & F & p & $F_{crit}$ \\ \hline 
Between groups & 297.253 & 1 & 297.253 &136.24& 1.11e-12 & 4.171 \\
Error (within group) &65.454& 30 & 2.388 & & \\
Total & 362.707& 31 \\ 
\hline \\
Results of sampled overall grade data. \\
\hline
Source of Variation & SS & df & MS & F & p & $F_{crit}$ \\ \hline 
Between groups & 127.932 & 1 & 127.932 & 81.27& 4.84e-12 & 4.171 \\
Error (within group) &47.226 & 30 & 1.574 & & \\
Total & 175.158 & 31 \\ \hline
\end{tabular}
\caption[]{Results of the one-way ANOVA tests.}
\end{center}
\end{table}

Since $p \leq 3.15e-07$ ($F>F_{\text{crit}}$) for all the ANOVA tests reported in Table 3, $H_0$ is rejected for the randomized assignment average ($p=1.36e-08<0.05$), midterm mark ($p=3.15e-07<0.05$), final mark ($p=1.11e-12<0.05$) and overall grade ($p=4.84e-12<0.05$) data. Hence, the differences in the means of the two populations are statistically significant. This result also agrees with the t-test results for each pair of means (which can be found in Appendix A). 

A two-way ANOVA test was also performed for all the random EoL and weekly assignment group data. This test is useful when data can be classified along two different dimensions. One dimension will be the different randomized groups, i.e., the random weekly and EoL assignments groups; the other will be the different ways the students were evaluated, i.e., the assignment, midterm, final and overall marks for each group. The two-way ANOVA test will allow us to test with 95$\%$ confidence the three following hypotheses:
\\ 
\begin{itemize}
\item[$H_{0_1}:$] there is no statistically significant difference between the population means of the two different groups,
\item[$H_{0_2}:$] there is no statistically significant difference in the effectiveness of how the populations were taught or evaluated, and their performance in the class, 
\item[$H_{0_3}:$] there is no statistically significant difference between the interaction of the groups and performance/evaluation. \\
\end{itemize}

The results of the two-way ANOVA test can be seen in Table 4. Since the p-value corresponding to the different groups (EoL and weekly) is $p=6.05e-26<0.05$, the null hypothesis $H_{0_1}$ is rejected. This implies that the different groups of students are statistically different. The p-value corresponding to the different types of course evaluation (i.e., assignments, midterm, final and overall grade) is $p=5.26e-12<0.05$, and so the null hypothesis $H_{0_2}$ is also rejected. Hence, there is a statistically significant difference between the effectiveness of how the different groups where taught or evaluated (or how the course was structured), and the outcome (i.e., marks) of the two groups. The two-way ANOVA test cannot say what the difference is, but whatever the difference may be, it significantly effected the performance or understanding of the two different groups of students. 

\begin{table}[h]
\begin{center}
\begin{tabular}{ lc c c c c c c c}
\hline
Source of Variation & SS & df & MS & F & p & $F_{crit}$ \\ \hline 
Group & 143.437 & 1 & 143.437 & 58.633& 5.36e-12 &3.920\\
Evaluation types &504.637 & 3&168.212 & 68.761 & 6.05e-26 & 2.680\\
Interaction & 599.861 & 3 &199.954 & 81.736 &7.21e-29 &2.680\\ 
Error & 293.561 & 120 &2.446 &\\
Total & 1541.495 &127 \\ \hline
\end{tabular}
\caption[]{Results of the two-way ANOVA test for random overall grade data.}
\end{center}
\end{table}

The p-value of the interactions is $p=7.21e-29<0.05$ and so, the null hypothesis $H_{0_3}$ is rejected. We conclude there are statistically significant differences in the interaction between the different types of course evaluation (i.e., assignment, midterm, final and overall marks) and the groups of students. We can plot the mean interactions to examine the interactions more carefully. These plots can be seen below in Fig. 10. Lines that are roughly parallel are indications of the lack of interaction, while lines that are not roughly parallel indicate interaction.

\begin{figure}[h]
\centering
\subfigure[Interaction factor 1.]{ 
\includegraphics[scale=0.44]{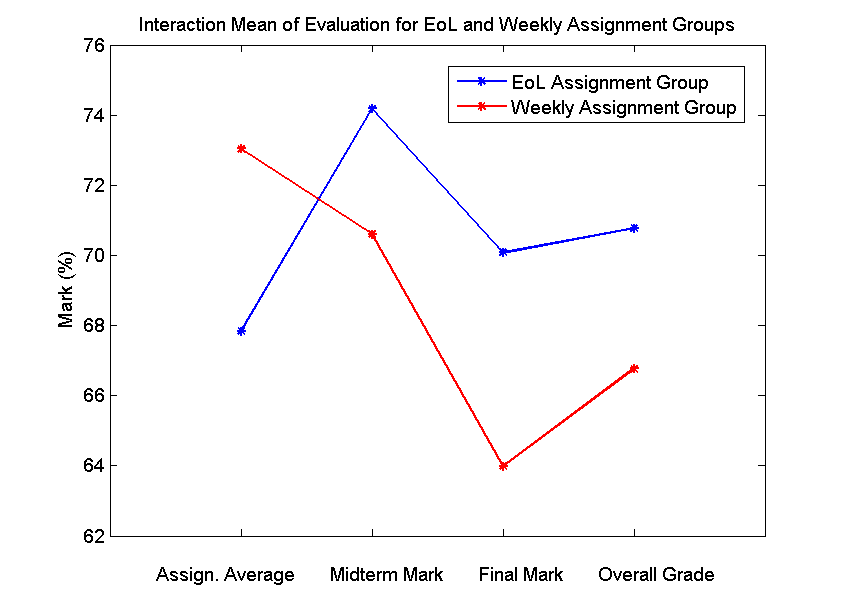}}
\subfigure[Interaction factor 2.]{ 
\includegraphics[scale=0.41]{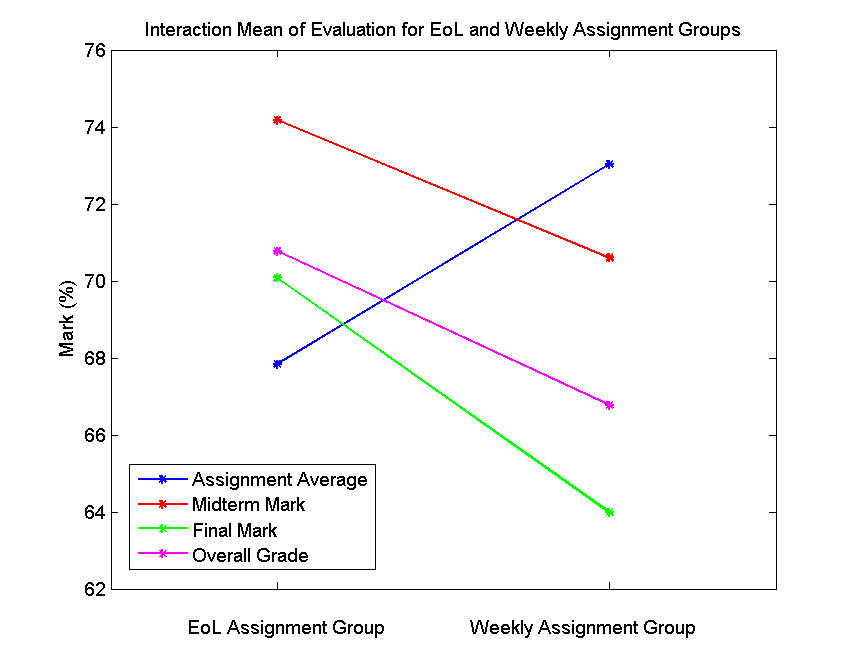}}
\caption[]{The mean interactions between the levels of the two factors}
\end{figure}

Therefore, in summary, these are incredibly small p-values indicating that the null hypotheses do not seem appropriate. This strongly indicates that the groups of students were not comparable. It appears there was a fundamental difference between how the students from the weekly assignment group and EoL assignment group absorbed and understood the MATH 235 material. However, in order to make any further statements, more data would have to be collected and analyzed.

\noindent \subsection{Conclusions for the restructuring of assignments}

In conclusion, by comparing Figs. 5 to 10 and the obtained p-values for the ANOVA tests, it appears that there is a statistically significant difference between the performance of the weekly assignment group and EoL assignment group. In addition, the results depicted in Figs. 7 to 9 seem to support the concept illustrated in Ebbinghaus' forgetting curve. That is, if students are able to review and test their knowledge on a more frequent basis, they will better understand the material and thus, perform better in classes. Perhaps if the math assignments, at least for certain courses, were structured in a similar fashion as the EoL assignments rather the weekly assignments, the students may obtain higher grades. Furthermore, from the survey results, we can see that at least for a portion of the students, such a structure may increase confidence, reduce anxiety and stress.

\noindent \section{Other Considerations} 

After reviewing the survey results discussed above, I spent much time talking with students individually and in small groups. All the feedback inspired the suggestions that I will now present. As a note to the reader, one suggestion is implementing better ways to integrate technology into the education system. I think this is an important aspects that needs serious consideration, and so it will be the focus of the proceeding paper.

\noindent \subsection{Basic Mental Health Training and Fall Reading Week}

During orientation week training at the University of Waterloo, orientation leaders are told to expect one out of two students to have some sort of mental health issue. Orientation leaders are also told to be alert for suicidal students. This high value suggests that mental health training for professors would be beneficial. Recall that from the student surveys, out of 107 students, 72$\%$ think professors should have such training. Desperate, mentally vulnerable people have the potential to be a danger to themselves or others; the faculty needs to know how to deal with such situations.

A wide variety of Canadian universities and colleges have conducted surveys showing that poor mental health, including suicidal thoughts, attempts and overall rates, is becoming a larger issue \cite{8}. One proposal to counteract this pandemic is to introduce a fall reading week. This is a solution that could be easily implemented; and in fact, many colleges and universities have already done so. For instance, Flemming College, Brock University, Wilfrid Laurier University, and Queen's University all now have fall reading weeks \cite{9}. 

A recent survey at the University of Waterloo showed that 74$\%$ of over 6,000 students polled want a fall reading week. However, according to university spokesman Nick Manning: ``implementing a reading week at the University of Waterloo is complicated by the demands of our co-op education program. So it's not as straight forward as the other 14 universities in Ontario who have been able to implement a reading week" \cite{10}. Nonetheless, the University of Waterloo senate will be voting hopefully this year on the perspective of having a fall reading week \cite{11}. Although, this was done approximately ten years ago, and the referendum never passed. A fall reading week at the University of Waterloo would only require two additional school days to be added to the fall term \cite{10}.

\noindent \subsection{Mental Health Days}

Another suggestion would be for the university to consider students' mental health in the same fashion as physical health, i.e., universities and colleges could employ mental health days like we do sick days. For instance, if a student is physically ill, especially with the influenza virus, alternative arranges or exceptions for courses are usually made for the student. Until recently, in the case of influenza, the University of Waterloo created a special form when H1N1 was at its peak. Students could submit the `declaration of absence for influenza-like illness' form online excusing them from class and the corresponding missed work without being penalized \cite{16}. Consider if a student for instance was going through a suicidal episode, a similar form could be equally beneficial, humane and perhaps even reduce drop-out rates \cite{17}. Upon surveying my students I found that:

\begin{itemize}
\item 71$\%$ of 102 students said: \textit{there should be mental health days like there are sick days.}

\item 67$\%$ of 95 students said: \textit{I think mental health days would be useful in a university setting.}
\end{itemize}

I personally hypothesize that mental illness is like the seasonal flu - anyone can have it or contract it. If this conjecture turns out to have enough supporting evidence, implementing mental health days especially seems appropriate. Interestingly, in a survey, students were asked if they agreed with such a hypothesis and 76$\%$ of 45 students agreed.

\begin{figure} [ht]
\centering
\includegraphics[scale=0.54]{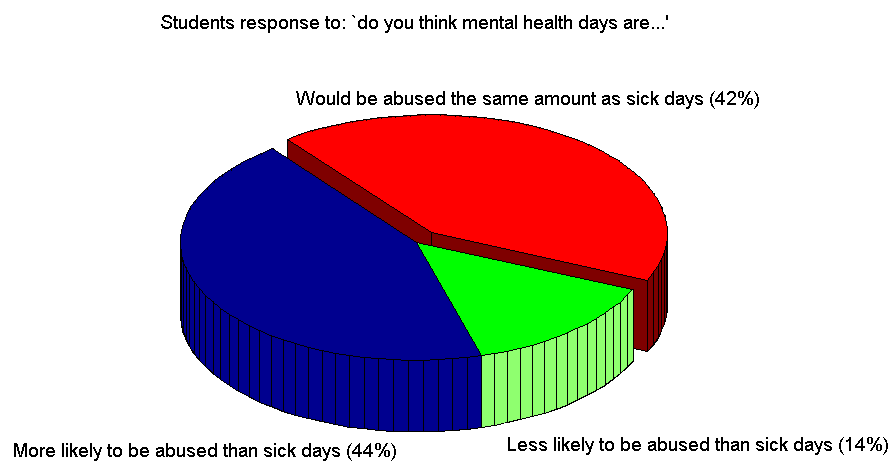}
\caption[]{Distribution of student responses (111 answered this question).}
\label{fi:1}
\end{figure}

A typical concern regarding the implementation mental health days is that they will be abused far more than sick days. Thus, in a survey, I asked my students what their view were on the potential abuse of mental health days. Fig. 11 shows the distribution of the 111 students who responded. One can see that 44$\%$ of the students said mental health days are more likely to be abused than sick days; whereas 42$\%$ did not think there would be a difference. My hypothesis is that mental health days would be abused the same amount as sick days. As some students have pointed out in the survey: 
\begin{itemize}
\item \textit{They would be abused the same amount as sick days, in some cases, due to the physical effects of certain mental illnesses, mental health days $==$ sick days.}
\item \textit{I think mental health is the same as ``sick", both can make it difficult to perform at your best.'}
\end{itemize}

\noindent\subsection{A Hypothesis of Mine - The Importance of Socialization}

One complaint that I often hear from students (and agree with) is that ``many mathematicians are socially-inept." The comparison of STEM individuals to those of autism-spectrum disorder (e.g., Asperger syndrome) is not uncommon, especially by the student body. Typically, students label and associate `social-ineptness' when their peers or professors seem to lack empathy, have abnormal social interaction(s), minimal social interactions(s), or scant communication(s) \cite{20}. Although this may initially sound offensive, it is supported by modern research. 

For example, one study was conducted in 2001 with a group of 840 students from Cambridge University (454 males, 386 females) where each student was given an autism-spectrum quotient test (AQ) \cite{AQ}. This test attempts to measure autistic traits for `individuals of normal intelligence.' The researchers found that scientists obtained higher AQ scores compared to non-scientists. In particular, it was found that within scientific fields, mathematicians had the highest AQ scores. Physical scientists, computer scientists and engineers also had higher scores compared to students in `human or life-centered' sciences (specifically medicine, veterinary science and biology) \cite{AQ}. These results also agreed with an earlier study carried out in 1998 again at Cambridge University. It was found that autism occurred ``significantly more often in families of students in the fields of physics, engineering, and mathematics'' \cite{AM}.  Overall, ``it seems that people who have autistic traits may find it easier or more natural to work in these fields," said Simon Baron-Cohen; the co-director of the Autism Research Centre at Cambridge University \cite{21}.

I found these studies to be quite intriguing since it has also been revealed that on average, millennials spend around seven to eleven hours a day in front of some sort of computer screen (desktop, tablet, smart phone etc.) \cite{22}. This indicates that the younger generations are not socializing (face-to-face) nearly as much as previous generations. Hence, children now do not gain as much experience physically socializing with others, which can be quite problematic. Learning how to read body language, especially facial expressions, is an imperative part of physiological development \cite{23}. Facial expressions for instance are a form of nonverbal communication that conveys emotion.  In fact, facial expressions are one of the primary ways that emotion is communicated between individuals \cite{24}. Being able to witness a person's reaction activates mirror neurons, allowing one to simulate what another is experiencing \cite{25}. 

A part of the brain that plays a key role in facial recognition is the amygdala \cite{28}. As we currently understand, the amygdala which is part of the limbic system in the brain, processes emotional reactions, memory formation and decision making \cite{26}. It plays an essential role in storing and forming memories associated with emotional experiences \cite{27}, and researchers speculate that the amygdala may be highly involved in recognizing negative emotions \cite{29}. Individuals with a larger amygdala are better able to recognize the emotional state of others through facial expressions. Therefore, researchers hypothesize that larger amygdalae may be an indication of greater emotional intelligence. The development of the amygdala is correlated with involvement in social networks. More specifically, the size of the amygdala increases according to the number of social interactions a person has. The variations and complexity of different social groups also influences development \cite{30}, \cite{31}. 

In addition, several studies have indicated that the amygdala has a significant role in mental states and hence, poor mental health \cite{32}. For instance, some studies have shown that children with anxiety disorders usually have a smaller left amygdale. An overactive left amygdala seems to be associated with borderline personality disorder and depression; whereas an underactive left amygdala seems to be correlated with psychopathy. In general, the amygdala seems to be connected with social anxiety, obsessive compulsive disorders, social phobias, bipolar disorder and not surprisingly, autism-spectrum disorders \cite{32}, \cite{33}, \cite{34}, \cite{35}. Studies have demonstrated that people with autism-spectrum disorders have a difficult time identifying facial expressions. In particular, some research has suggested that such disorders escalate due to a lack of physical social experience: ``the development of face perception and social cognitive skills are supported by the amygdala–fusiform system, and that deficits in this network are instrumental in causing autism'' \cite{36}.

Due to these studies and findings, I hypothesize that all mathematicians are ``a little bit autistic" and potentially, more susceptible to poor mental health. Consider that to be a successful mathematics student, the student has to spend more time alone doing problems compared to individuals studying less technical fields. To complete an undergraduate degree in mathematics, at least four years has to be dedicated to learning. Typically such an endeavour starts around the age of 17 - 19, sometimes even younger. If a student then enrolls into a masters or doctorate degree, that is another two to six years of intensive work. In total (not including post-doctoral fellowships), that is approximately four to ten years during the period in one's life where individuals grow into the people they want to be \cite{Jay}. This is a lot of time being spent alone. This may imply that the parts in the brain responsible for empathy and socialization, such as the amygdala, will not sufficiently develop; especially compared to people with more complex social networks and interactions. 

If this hypothesis is true, we may have a serious problem. If a STEM background is necessary in the twenty-first century, a time where computers and the Internet are an integral part of our lives, will people start becoming less socially aware and less empathetic (for some more information see \cite{37})? Could the people who are needed to solve the world's biggest problems be more likely to suffer from psychological disorders? These are questions that we cannot afford to ignore or avoid.

This hypothesis also seems to be connected with a major complaint people, and especially women have regarding mathematics or generally STEM education: it is lonely. Social interaction is more important biologically for women than it is for men \cite{38}, and women are more likely to change academic and professional paths than men due to their lack of confidence \cite{CBC}. Perhaps these are reasons why there are more men in fields like maths, physics and engineering. A potential solution may sound simple: make mathematics and the education system more social with the help of technology (which will be the focus of the next paper). In my opinion, we need to create an open, positive, social learning environment to nurture healthy educated people.

\noindent\section{Conclusions}

Mathematics is the only subject where proofs exist. As Chris Rock said: ``there is mathematics and everything else is debatable.'' It is a beautiful subject that is based in truth without ambiguity. However, to learn mathematics is no trivial task. It requires an excellent teacher, and a good state of mental health so the mind's computational power can be allotted to learning. How best to learn and teach mathematics is debatable and generally is gaining more attention by researchers. Nonetheless, it is essential to listen to the students' needs and suggestions. Providing a positive, open and safe learning environment is necessary. In the post-secondary education system, this requires a lot of time and effort from the university and in particular, its faculty. However, that is the purpose of post-secondary institutions: to be a place of higher learning, open-mindedness and innovation. We need to actively prepare individuals to create and maintain a dynamic, functional, altruistic society that has many challenging problems to solve within the next 50 - 100 years \cite{4}. Without awareness of mental health, this goal is very unlikely to be achieved.

\noindent\section{Acknowledgments}

I would like to thank all of the students who participated in these surveys. Thanks for your openness and honesty. Also, I would like to thank the following undergraduate math students for their discussions, ideas, questions, interpretations and suggestions: Amar Bhalla, Andre Jodat-Danbrani, Andrew Park, Ashna Singh, Glen Chalatov, Ivan Li, Jessie Huang, Jessie Yeung, Joshua Thompson, Cheyenne Guo, Varmanaa, Mohammad Ali, Monica Pierce and all others who shared their stories.

\newpage

\noindent\section{Appendix A: An exhaustive statistical analysis of the weekly and EoL assignment group data}

To compare the two different groups, several statistical tests were performed. F-tests were first carried out to establish if the difference in variances for the populations is statistically significant. Then, two-sample student's t-tests were used to determine if there is a statistically significant difference in means. The assumption on whether to perform t-tests with equal or unequal variances was determined by the results of the F-tests. A one and two-sample z-test for means was also performed. The one-sample z-test was carried out to determine if the samples come from a normal distribution with a specified mean and standard deviation. The two-sample z-test was performed to determine if there is a statistically significant difference between the means with known variances. All tests performed where done separately for the two groups' assignment average, midterm mark, final mark and overall grade data. For each test, the null hypothesis, denoted by $H_0$, and the corresponding result with 5$\%$ significance level, are given below. The independent sample populations for all tests are the weekly assignment and EoL assignment groups' data.

\noindent \subsection{Two-sample F-test for variances}

The two-sample F-test will be used to test if the variances of the two populations: the weekly and EoL assignment groups, have a statistically significant difference with 95$\%$ confidence. We define the null hypothesis as follows:

\begin{itemize}
\item[$H_0$:] the two independent samples come from distributions with equal variances, against the alternative that the variances are not equal in the underlying distributions.
\end{itemize}

The F-test returns a test decision for $H_0$; the corresponding p-value, denoted by p; the test statistic for the F-test, denoted by F; as well as the lower and upper boundaries of the 95$\%$ confidence interval for the true variance ratio. These results can be seen in Table V. The test statistic is a ratio of the two sample variances. In particular, it is defined as $$F=\frac{s_1^2}{s_2^2},$$

\noindent where $s_1$ and $s_2$ are the sample standard deviations of the EoL and weekly assignment group, respectively. The more the ratio deviates from one, the more likely $H_0$ will be rejected ($F_{\text{critical}}=0.85183$).

The returned results indicate that the null hypothesis cannot be rejected for the assignment average ($p=0.2580>0.05$) and midterm mark ($p=0.1202>0.05$) data. However, the null hypothesis is rejected for both the final mark ($p=0.004<0.05$) and overall grade ($p=0.0026<0.05$) data. In other words, the two assignment average and midterm mark samples come from distributions where the difference in the variances is not statistically significant. However, the variance difference for the final mark and overall grades samples are statistically significant.

\begin{table}[h]
\begin{center}
\begin{tabular}{ c c c c c}
\hline
Data Type & F-stat & p-value & CI (95 $\%$)&Null Hypothesis\\ \hline 
Assign. Average & 0.8956 & 0.2580&[0.7386, 1.0842 ]& Cannot reject \\ \\
Midterm Mark & 0.8594 & 0.1202 &[0.7088, 1.0404]& Cannot reject \\ \\
Final Mark & 0.7573 & 0.0044& [0.6245, 0.9167]&Reject\\ \\
Overall Grade & 0.7452 &0.0026&[0.6146, 0.9021]& Reject \\ \hline \\
\end{tabular}
\caption[]{Results of the two-sample F-test for variances of the marks composition data set.}
\end{center}
\end{table}

\noindent \subsection{Two-sample t-test for means}

The two-sample t-test will be used to test the equality of the population means of the weekly and EoL assignment groups with 95$\%$ confidence. However, from the F-test, two separate t-tests must be performed: one assuming that the population variances are equal (for the assignment average and midterm mark data), and another assuming that the population variances are not equal (for the final mark and overall grade data). Therefore, we define the two null hypotheses as follows:

\begin{itemize}
\item[$H_{0_1}$:] two independent random samples come from distributions with equal means and equal, unknown variances, against the alternative that the data comes from populations with unequal means.
\item[$H_{0_2}$:] two independent random samples comes from distributions with equal means but unknown, unequal variances, against the alternative that the data comes from populations with unequal means (this is referred to as a heteroscedastic t-test).
\end{itemize}

\begin{table}[h]
\begin{center}
\begin{tabular}{ c c c c c}
\hline
Data Type & t-stat & p-value & CI (95 $\%$)&Null Hypothesis\\ \hline 
Assign. Average & $t_1=-3.2254$& 0.0013 &[-7.8009, -1.8984]& $H_{0_1}$: Reject \\ \\
Midterm Mark & $t_1=3.1124$ & 0.0019&[1.5023, 6.6322]& $H_{0_1}$: Reject \\ \\
Final Mark & $t_2=8.3862$ & 0.0000& [8.3862, 13.4869]& $H_{0_2}$: Reject\\ \\
Overall Mark & $t_2=4.0744$ &0.0001&[2.4649, 7.0486]& $H_{0_2}$: Reject \\ \hline \\
\end{tabular}
\caption[]{Results of two-sample t-test for means assuming equal, unknown variances for the assignment average and midterm mark data; and the results of the two-sample t-test for means assuming unequal, unknown variances for final mark and overall grade data.}
\end{center}
\end{table}

The t-test returns a test decision for $H_0$; the corresponding p-value, denoted by p; the test statistic for the t-test, denoted by t; as well as the lower and upper boundaries of the 95$\%$ confidence interval for the difference in population means. These results can be seen in Table VI. The test statistic is defined as follows:

$$ t_1= \sqrt{\frac{(n-1)s_1^2+(m-1)s_2^2}{(n+m)-2}}$$
$$ t_2= \frac{m_1-m_2}{\sqrt{\frac{s_1^2}{n}+\frac{s_2^2}{m}}}$$

\noindent where $t_1$ ($t_{\text{critical}}=1.962782$) and $t_2$ ($t_{\text{critical}}=1.963054$) are the test statistics corresponding to the t-test with equal and unequal variances, respectively; $s_1$ and $s_2$ are the sample standard deviations; $m_1$ and $m_2$ are the sample means; and n and m are the sample sizes of the EoL and weekly assignment group, respectively.

The returned results indicate that both null hypotheses are rejected in all cases with 5$\%$ significance level. In other words, the two assignment average ($p=0.0013<0.05$), midterm mark ($p=0.0019<0.05$), final mark ($p=0.0000<0.05$), and overall grade ($p=0.0001<0.05$) samples come from distributions with where the difference in means are statistically significant.

\noindent \subsection{One-sample z-test for mean}

The one-sample z-test will be used to determine with 95$\%$ confidence if each sample comes from a normal distribution with a specified mean, denoted by $\mu$, and standard deviation, denoted by $\sigma$. Note that these values for both groups can be found in Tables 1 and 2. The test was carried out separately on the EoL and weekly assignment groups for each of the assignment average, midterm mark, final mark and overall grade data. Thus, the null hypothesis is defined as:
\begin{itemize}
\item[$H_0$:] the sample comes from a normal distribution with mean $\mu$ and standard deviation $\sigma$, against the alternative that the mean is not $\mu$.
\end{itemize}

The one-sample z-test returns a test decision for $H_0$; the corresponding p-value, denoted by p; the test statistic for the z-test, denoted by z; as well as the lower and upper boundaries of the 95$\%$ confidence interval for the true population mean. These results can be seen in Tables VII and VIII. The test statistic is defined as follows:
$$z=\frac{\bar{x}-\mu}{\frac{\sigma}{\sqrt{n}}} $$

\noindent where $\bar{x}$ is the sample mean, $\mu$ is the population mean, $\sigma$ is the population standard deviation, and n is the sample size.

\begin{table}[h]
\begin{center}
\begin{tabular}{ c c c c c}
\hline
Data Type & z-stat & p-value & CI (95$\%$)&Null Hypothesis\\ \hline 
Assign. Average & 0.0741 & 0.9410 &[66.1514, 69.9937]& Cannot reject \\ \\
Midterm Mark & -0.6353& 0.5252&[72.8195, 76. 1129]& Cannot reject \\ \\
Final Mark & -0.2334 & 0.8155& [69.2596, 72.3700]& Cannot reject\\ \\
Overall Mark & 0.4518 &0.6514 &[69.9441, 72.6886]& Cannot reject \\ \hline \\
\end{tabular}
\caption[]{Results of z-test for the EoL assignment group.}
\end{center}
\end{table}
\begin{table}[h]
\begin{center}
\begin{tabular}{ c c c c c}
\hline
Data Type & z-stat & p-value & CI (95$\%$)&Null Hypothesis\\ \hline 
Assign. Average & -0.0625 & 0.9502 &[70.7353, 75.1248]& Cannot reject \\ \\
Midterm Mark &0.3919 & 0.6951 &[68.4038, 72.3942]& Cannot reject \\ \\
Final Mark & -0.1196 & 0.9048& [57.8830, 61.8734]& Cannot reject\\ \\
Overall Mark &-0.4807 & 0.6307 &[64.7639, 68.3553]& Cannot reject \\ \hline \\
\end{tabular}
\caption[]{Results of z-test for the weekly assignment group.}
\end{center}
\end{table}

The results indicate that the null hypothesis for all tests cannot be rejected since the p-value in each case is greater than 0.05. Therefore, each sample comes from a normal distribution with the specified mean and standard deviation from Tables 1 and 2. 

\noindent \subsection{Two-sample z-test for means with known variances}

The two-sample z-test will now be performed to determine if there is a statistically significant difference between the two population means (for the marks composition data set) with 95$\%$ confidence. The known mean difference between the EoL and weekly assignment group is considered in the test. In particular, the mean difference for the assignment average, midterm mark, final mark and overall grade data is 4, 4, 10 and 5, respectively. The null hypothesis is defined as follows:
\begin{itemize}
\item[$H_0$:] that there is no difference between two population means with known variances, against the alternative that there is a statistically significant difference.
\end{itemize}

The two-sample z-test returns a test decision for $H_0$; the corresponding p-value, denoted by p; and the test statistic for the z-test, denoted by z; as well as the critical z-value. These results can be seen in Table IX. 

\begin{table}[h]
\begin{center}
\begin{tabular}{ c c c c c}
\hline
Data Type & z-stat & p-value & z-critical &Null Hypothesis\\ \hline 
Assign. Average & 0.5677 & 0.5702 &1.95996& Cannot reject \\ \\
Midterm Mark & -6.1368 & 8.42e-10 &1.95996& Reject \\ \\
Final Mark & -16.12 & 0 &1.95996& Reject\\ \\
Overall Mark & -8.3607 &0 & 1.95996 & Reject \\ \hline \\
\end{tabular}
\caption[]{Results of the two-sample z-test with hypothesized mean difference of 4, 4, 10 and 5 for the assignment average, midterm mark, final mark and overall grade data, respectively.}
\end{center}
\end{table}

The returned results indicate that the null hypothesis cannot be rejected for the assignment averaged data ($p=0.5702>0.05$) with 5$\%$ significance level. However, the null hypothesis is rejected for the midterm mark ($p=8.42e-10<0.05$), final mark ($p=0<0.05$) and overall grade ($p=0<0.05$) data. In other words, with the exception of the assignment average data for the EoL and weekly assignment group, there is a statistically significant difference in the means of the two groups.

\newpage

\noindent\section{Appendix B: Additional comments from students}

\noindent \subsection{On mental health and university life in general}
\begin{itemize}


\item \textit{Before Waterloo, I didn't have anxiety and depression. I feel very alone here, like very few people with the power to help actually care enough about me to do so.}

\item \textit{I definitely think that university has put a lot more stress on me. I feel like from about the second or third week of classes right until my last exam that I'm constantly stressed. I feel like there's always something I should be doing, whether it's an assignment, textbook readings, going over notes, or studying for a test. I find it's hard for me to have a balance between socializing and school because I know that I'm in school for a reason so I want to do everything I can to succeed, leaving little free time left over for myself.}

\item \textit{In high school I was successful and worry free, but when I came to university, all my friends and I ever talk about is grades and homework making it a very stressful environment for me. When I don't do as well as my friends, I feel dumb and like a failure. }

\item \textit{The amount of stress placed upon students tends to have a negative effect on their mental health, however I have found that the large student body has also given me many opportunities to connect with others in similar situations and help myself deal with the stress.}

\item \textit{Sometimes profs is this [math] faculty can really put a student down. I remember once I failed a course and I was really concerned about it. I went to an advisor to seek for some help and comfort and the advisor straight up told me I shouldn't be in this program because I'm clearly very bad at math and had really bad marks. It made me so much worse. He could have chosen better words.}

\item \textit{I think that the university needs to implement a reading week in fall semester. I'm not just saying this because I WANT time off of school, but I think we NEED time off of school. By the time Thanksgiving comes around, the stress of midterms has already started. I think having a week off would definitely improve my mental health, giving me time to relax or even time for me to catch up on work, which I know for a fact would greatly decrease my stress. The reading week in February helps decrease my stress levels a lot.}

\item \textit{Instead of having someone come in and state `councelling services is here, AccessAbility is here' maybe have someone's testimony on how they were helped and what caused them to need to seek help in the first place. I came to Waterloo with undiagnosed major depressive disorder, generalized anxiety disorder and ADHD combined. As a result, I felt very let down as there was too little awareness of these services and how they actually do help students. After nearly failing out of my first program, seeking immense help, medication, gaining access to tutors and extensions, and getting diagnoses, I am finally getting back on my feet; I feel like I was greatly let down by my faculties for their poor ability to create awareness and very poor attempts to dismiss stigma. TAs, ISA and whatnot should have to take courses in how to approach students with mental health problems. Having to take a mental health day should not be treated as someone just merely being sick and missing a class. It should be made known that note takers are available and paid note takers (trained by the faculties) should be made available to students in need. }

\end{itemize}

\noindent \subsection{On EoL assignments}

\begin{itemize}

\item \textit{I don't enjoy having more assignments, however I feel they are helping me understand content better.}

\item \textit{I like that they spread the work of assignments over the entire week, rather than concentrate them on one night. Plus if I ever miss one, I'm not as scared as missing a weekly assignment.}

\item \textit{At first I was like ``what the hell is this?'' but after a few of them I started to realize that they're intended to help me, not piss me off. I think that I've been keeping up with the material much better with 3 mini-assignments per week as opposed to the 1 big assignment from 136. I feel much more confident in class as I feel that I'm able to understand more material.}

\item \textit{I don't like EoL assignments because I have less flexibility to schedule when to do them (I can only choose between a 2-day period instead of a 1-week period). When the term gets busier, I will be much more likely to skip these assignments since I cannot do them a few days later.}

\item \textit{They're great - they're actually making me review the material. My thought process for the EoL assignments has shrunk from the normal assignment times of 30 minutes to a mere 10 minutes. This forces me to recall the knowledge, and I enjoy being able to finish them at a quicker pace. However, I do feel slightly stressed that they're due 3 times a week. Perhaps the frequency can be changed to twice a week - I will feel at ease $:)$}

\item \textit{I think the EoLs are very useful and wish my other classes would also do them.}

\item \textit{They're great. Reviewing my notes after class doesn't test me the way EoLs do. Weekly assignments seem like a hassle as they are long and often we haven't learned all material required for the assignments in time. EoLs don't seem like a hassle and are mostly relevant to what we learned the day of or recently and thus are a great way of learning and reviewing class notes. Please don't get rid of them!}

\item \textit{The end of lecture assignments make me keep up with the material, but I just have to figure out how to manage my time better.}

\item \textit{At first I thought the EoL assignments would be too much work, but I actually really like them! I find they help me actually understand the material from every class, and if I don't understand the material right away, I learn it sooner than I would have with weekly assignments. The weekly assignments in previous math classes (135, 136, 137 and 138) usually seemed difficult and quite daunting to me, but these are not so bad. Yes, there are some questions I don't always get correct or know exactly how to tackle, but having just learned the material is helpful in that it is fresh in my mind. Also, I find that having the assignments after every lecture makes me actually go over the notes from each lecture, which I will admit I did not do last year at all. Last year I would get to the weekly assignment and realize I hadn't really understood a lot from the week's lectures, and have to go back and try to understand it then. So I think this will really help me stay on task and hopefully do better in the course. So far, I think these assignments are great, actually!}


\item \textit{The EoLs have grown on me. They're a great way to ensure that I keep up with the course material because I want to do well on the assignments. Since I'm keeping up more with the content during the week this makes the lectures more effective and easier to understand. The one thing I am concerned about is once midterms start, whether or not I'll have enough time…}


\item \textit{At the start of the course when I heard about EoL assignments I thought it was going to be so bad and time consuming. I thought it would be so hard to keep on top of them and balance it out with my other courses. But now I'm actually surprised at how much better they are than doing weekly assignments where I would try to learn the entire week's material just before doing the assignment and it would take me much longer. I definitely think the EoL assignments are a great idea! $:)$}

\item \textit{Although they were a little tedious, I felt they did help with learning the new material in lecture and a way to learn as we went along in class.}

\item \textit{I didn't like them originally because I found that they were taking a lot of time and so I didn't feel as though I was putting enough time into my other courses. However, now I'm not sure that I spend more time on overall on them than I would on assignments. So now I'm kind of sitting on the fence as to whether I like them or not. I think they are good in making me review my notes and making sure that I understand the concepts as we go along as long as it's not to the detriment of my other courses.}

\item \textit{They help me stay on top of the material and they reinforce the material very well. I am better prepared for lectures because of these assignments. $:)$}


\item \textit{There's no way I could do 3 assignments a week for more than one class on top of the normal university work load. However, I think they're great for this class since the material is pretty complex and they've made me excited to do well this term. $:)$}

\item \textit{I think they are a really great idea! They help me stay on top of the material and retain what we are learning. They also provide a lot of questions to look back on when studying for midterm/final. }

\item \textit{$:)$ keeps me going to school at least}

\item \textit{I really like the idea of EoL assignment, it is helping me to keep up what I have just learned by forcing me to do these exercise. It divides the study time of one subject into many little sections throughout the week, and this is very efficient for absorbing the concepts. EoL is great! EoL rules!}

\item \textit{I like the fact that the EoL assignments are shorter but due more regularly than the average weekly assignment from other classes. It helps me to keep on top of the material rather than procrastinating until the day before it's due (not that I'm inclined to do that for the most part). I like them, but I think if I had them in every class I would feel overwhelmed so a part of the reason they're affective is because my other classes only have weekly assignments. Overall, a very bright idea for MATH 235.}

\item \textit{EoLs are a good idea, but sometime it might be too much for students who have a lot to do during the week, so online submission would be an awesome idea.}

\item \textit{Love em. The assignments keep me on top of my game.}


\item \textit{I believe that the EoL assignments were extremely helpful and I think that it would be very beneficial to the students and the university if they were incorporated into all of the courses in the math faculty. They will raise students marks and improve their understanding of the material. With these EoLs, profs will be able to cover more material and push students further so that they can achieve at a higher level.}

\item \textit{The EoL assignments were much easier to manage than 1 weekly assignment. It also allowed me to review the notes after every lecture which definitely helped me to understand the content better. (if only other courses had it too $:P$)}

\item \textit{Tedious but served their purpose if you kept up with them.}


\item \textit{I just think they should be less frequent, i.e., every other class.}


\item Are you less anxious or stressed coming to class after doing an EoL assignment? 

\begin{itemize}
\item \textit{Despite most likely having an anxiety disorder, lectures are one of the few things in my day that cause me no anxiety. I find that I can follow the lectures and in most cases can finish the examples on my own, which makes it a stress-free time.}

\item \textit{More anxious. I like being flexible with my time. That is, I like to spend a day focusing on 1-3 things. With EoLs, it forces me to know the material even when I don't feel ready for it. I know I can learn it all on one day of the week on my own anyway.}

\item \textit{I feel the least amount of stress in this class and I feel that I also have the highest level of understanding compared to my other courses.}
\item \textit{They are good but they add to the stress of finishing work to hand in. }

\item \textit{They are really helping decrease my stress level. I feel more comfortable with the material and I feel like they will allow me to be successful in the course.}

\item \textit{I generally liked them, though they were stressful.}

\item \textit{They're the best! I'm forced to keep up with the course content, and for the first time, I'm actually keeping up with the lectures instead of furiously copying down notes and being lost in the lectures. I also love that I'm not so stressed out -- I'm a procrastinator, and with the weekly assignments, I'd find myself doing everything the night before, often having to pull all-nighters. This would ruin my entire week and cause so much stress. I honestly think that this should be implemented in the calculus classes as well, and maybe even in some of the first year courses. }

\end{itemize}

\end{itemize}

\end{document}